# A possible solution to the solar neutrino problem: Relativistic corrections to the Maxwellian velocity distribution

Jian-Miin Liu*

Department of Physics, Nanjing University

Nanjing, The People's Republic of China

*On leave. Present mailing address: P.O.Box 1486, Kingston, RI 02881, USA.

Abstract

The relativistic corrections to the Maxwellian velocity distribution are needed for standard solar models. Relativistic equilibrium velocity distribution, if adopted in standard solar models, will lower solar neutrino fluxes and change solar neutrino energy spectra but keep solar sound speeds. It is possibly a solution to the solar neutrino problem.

All standard solar models [1,2] use the Maxwellian velocity distribution to complete the thermal averages of quantity $v\sigma$ estimating the rates of solar nuclear fusion reactions, where $v$ is the magnitudes of the relative velocities of interacting particles and $\sigma$ is their cross section. The Maxwellian velocity distribution is obviously inconsistent with special relativity. Special relativity is a part of the laws of Nature. We need to make the relativistic corrections to the Maxwellian velocity distribution for these well-constructed standard solar models. As the relativistic equilibrium velocity distribution, the corrected distribution is expected to be close-fitting to the Maxwellian distribution in the low-velocity part, to be substantially different from the Maxwellian distribution in the high-velocity part, and to vanish at the velocity of light.

One might argue that since no significant difference in the low-velocity part exists between the Maxwellian distribution and the corrected distribution, since most particles crowd in the low-velocity part even at the temperatures and densities in the Sun, there will be no significantly different results when we use whether this one or that one.

This argument is applicable only to the case that most relevant particles crowd in the low-velocity part and, concurrently, this low-velocity part is involved in statistical calculations. The calculations of solar sound speeds are of the case. The relevant particles are all solar ions. Most of them crowd in the low-velocity part and this part is involved in the calculations of sound speeds. However, the calculations of solar nuclear fusion reaction rates, and hence, of solar neutrino fluxes and solar neutrino energy spectra, do not pertain to the case. To create a nuclear fusion reaction, a proton or a nucleus must penetrate the repulsive Coulomb barrier and collides with another proton or nucleus. The height of Coulomb barrier is far above solar thermal energy: their ratio is typically greater than a thousand [2,3], so most solar protons and nuclei which crowd in the low-velocity part are forbidden to have a share in nuclear fusion reactions. Only few high-velocity or high-energy protons and nuclei can participate in nuclear fusion reactions. A difference in the high-velocity part between the Maxwellian velocity distribution and the relativistic equilibrium velocity distribution may cause significant results.

Velocity space is a space in which point pairs represent relative velocities. In the pre-relativistic mechanics, the velocity space is represented by

$$dY^2 = \delta_{rs} dy^r dy^s, \quad r,s=1,2,3, \tag{1}$$

in the usual velocity-coordinates $\{y^r\}$, $r=1,2,3$, where $y^r$ is the well-defined Newtonian velocity. This velocity space is characterized by (a) unboundedness and (b) the Galilean addition law.

In the special theory of relativity and the modified special relativity theory [4], the velocity space is described by

$$dY^2 = \delta_{rs} dy'^r dy'^s, \quad r,s=1,2,3, \tag{2}$$

in the so-called primed velocity-coordinates $\{y'^r\}$, $r=1,2,3$, [4-6] or by

$$dY^2 = H_{rs}(y) dy^r dy^s, \quad r,s=1,2,3, \tag{3a}$$

$$H_{rs}(y) = c^2 \delta^{rs}/(c^2-y^2) + c^2 y^r y^s/(c^2-y^2)^2, \quad \text{real } y^r \text{ and } y<c, \tag{3b}$$

in the usual velocity-coordinates, where $y=(y^r y^r)^{1/2}$, $r=1,2,3$, and c is the velocity of light [4-7]. The transformation connecting the primed and the usual velocity-coordinates is,

$$dy'^r = A^r_s(y) dy^s, \quad r,s=1,2,3, \tag{4a}$$

$$A^r_s(y) = \gamma \delta^{rs} + \gamma(\gamma-1) y^r y^s / y^2, \tag{4b}$$

with

$$\gamma = 1/(1-y^2/c^2)^{1/2},$$

and

$$y'^r = [\frac{c}{2y} \ln \frac{c+y}{c-y}] y^r, \quad r=1,2,3, \tag{5a}$$

$$y' = \frac{c}{2} \ln \frac{c+y}{c-y}, \tag{5b}$$

where $(y'^1, y'^2, y'^3)$ and $(y^1, y^2, y^3)$ represent the same point in the velocity space, and $y'=(y'^r y'^r)^{1/2}$, $r=1,2,3$ [5,6]. This velocity space is characterized by (a) and (b) in the primed velocity-coordinates, and also by (c) a finite boundary at c and (d) the Einstein velocity addition law in the usual velocity-coordinates. We call $y'^r$, $r=1,2,3$, the primed velocity. Its definition from the measurement point of view is given in Ref.[8]. The Galilean addition law of primed velocities links up with the Einstein addition law of corresponding Newtonian velocities [4,5].

The Euclidean structure of the velocity space in the primed velocity-coordinates convinces us of the Maxwellian velocity and velocity rate distributions in the primed velocity-coordinates, namely

$$P(y'^1, y'^2, y'^3) dy'^1 dy'^2 dy'^3 = N \left(\frac{m}{2\pi K_B T}\right)^{3/2} \exp\left[-\frac{m}{2 K_B T}(y')^2\right] dy'^1 dy'^2 dy'^3 \tag{6}$$

and

$$P(y') dy' = 4\pi N \left(\frac{m}{2\pi K_B T}\right)^{3/2} (y')^2 \exp\left[-\frac{m}{2 K_B T}(y')^2\right] dy', \tag{7}$$

where N is the number of particles, m their rest mass, T the temperature, and $K_B$ the Boltzmann constant. Using Eqs.(4a-4b) and (5a-5b) in Eqs.(6) and (7), we can find

$$P(y^1, y^2, y^3) dy^1 dy^2 dy^3 = N \frac{(m/2\pi K_B T)^{3/2}}{(1 - y^2/c^2)^2} \exp\left[-\frac{mc^2}{8 K_B T} \left(\ln \frac{c+y}{c-y}\right)^2\right] dy^1 dy^2 dy^3. \tag{8}$$

and

$$P(y) dy = \pi c^2 N \frac{(m/2\pi K_B T)^{3/2}}{(1 - y^2/c^2)} \left(\ln \frac{c+y}{c-y}\right)^2 \exp\left[-\frac{mc^2}{8 K_B T} \left(\ln \frac{c+y}{c-y}\right)^2\right] dy \tag{9}$$

as the relativistic equilibrium velocity and velocity rate distributions in the usual velocity-coordinates. $P(y^1, y^2, y^3) dy^1 dy^2 dy^3$ and $P(y) dy$ are so named because they are based on the velocity space in special relativity and the modified special relativity theory.

The relativistic equilibrium velocity and velocity rate distributions reduce to the Maxwellian velocity and velocity rate distributions for small velocities. They both contain a relatively depleted high-energy tail and vanish at the velocity of light [5,6].

Evidently, the nuclear fusion reaction rate based on the relativistic equilibrium velocity distribution, R, has a reduction factor with respect to that based on the Maxwellian velocity distribution, $R_M$ [9],

$$R = \frac{\tanh Q}{Q} R_M, \tag{10a}$$

$$Q = \left(2\pi z_1 z_2 \frac{K_B T}{\mu c^2} \frac{e^2}{\hbar c}\right)^{1/3}. \tag{10b}$$

$R_M$ is actually the first-order approximation of R. Since $0 < Q < \infty$, the reduction factor satisfies

$0 < \tanh Q/Q < 1$. That gives $0 < R < R_M$. The reduction factor depends on the temperature, reduced mass, and atomic numbers of the studied nuclear fusion reactions. The relativistic equilibrium velocity distribution, if adopted in standard solar models, will lower solar neutrino fluxes and change solar neutrino energy spectra but keep solar sound speeds. It is possibly a solution to the solar neutrino problem.

ACKNOWLEDGMENT

The author greatly appreciates the teachings of Prof. Wo-Te Shen. The author thanks Prof. Gerhard Muller and Dr. P. Rucker for suggestions.